\documentclass[12pt,english]{article}
\usepackage[latin1]{inputenc}
\usepackage{graphicx}
\usepackage{epsf}
\begin{document}

\title {"Antiferromagnetism" in social relations and Bonabeau model}

\author{\textbf{ G\'{e}rard Weisbuch$^1$ and Dietrich Stauffer$^2$} \\
1: Laboratoire de Physique Statistique, \footnote{Laboratoire associ{\'e}
au CNRS (URA 1306), {\`a} l'ENS et aux Univ. Paris 6 et Paris 7}\\
 Ecole Normale Sup\'{e}rieure, Paris, Euroland.\\
2: Institute for Theoretical Physics, \\
Cologne University, D-50923 K\"oln, Euroland.}

\maketitle


\abstract{ We here present a fixed agents version of an original
model of the emergence of hierarchies among social agents
first introduced by Bonabeau \textit{et al}. Having interactions occurring
on a social network rather than among 'walkers' doesn't
drastically alter the dynamics. But it makes social structures
more stable and give a clearer picture of the social
organisation in a `mixed' regime.}

\vskip 1cm

\section{Introduction}

  A number of models in socio- and econo-physics are inspired 
from magnetism, starting with the pioneering work of F\"ollmer (1974)
\cite{Follmer}.
 They concern the modeling of opinion dynamics and assume 
for instance binary opinions of agents distributed on some 
social networks, with variants which include the mean field
approximation. Agents adjust their opinion under the influence of their
neighbours. In many models, agents follow the opinion of their 
neighbours via for instance some "voting" scheme: an agent would then
check the opinions of its neighbours and adjust its own opinion
according to the majority's opinion. The observed dynamics resemble 
ferromagnetism.

  But a large number of other models 
start from the opposite assumption.

In the minority game \cite{minority}, applied to financial 
markets, agents choose to behave in opposition
with the majority: if the others are buying, you should sell
(and the opposite). The minority game is only one of such models to which one 
can relate the model of the emergence of social classes based on the ultimatum 
game of Axtell \textit{et al} \cite{Axtell}, and Bonabeau \textit{et al} 
model \cite{Bonabeau} of the emergence of social hierarchies. This Bonabeau 
\textit{et al} 
approach, followed by the contributions of Ben-Naim and Redner \cite{Ben-Naim}
(taking a mean field approach) and of Stauffer \cite{Stauffer}, is the original
inspiration for the work presented here.   

   We  consider agents $i$ whose internal
state $h(i)$ is adjusted on the occasion of binary encounters.
 $h(i)$ can be interpreted as strength and encounters as fights.
On the occasion of a fight, the winner, usually the stronger agent, wins
and its strength is increased by one. The looser strength is decreased
by one. By usually, we imply a probabilistic process with some ''thermal''
probability.  Bonabeau \textit{et al} original model is a walkers' model: agents
are moving across a square lattice, and interact only when they encounter 
on the same lattice site. The main parameter of these models is the $\beta$ 
discrimination parameter for the outcome of the fight process.
In Bonabeau \textit{et al} a phase transition is observed: at low  $\beta$ 
values, a ''disordered" phase is observed with a Gaussian distribution 
of agent strengths, while at high  $\beta$ values, a large fraction of agents 
 have extreme strength, positive or negative. Several variants yield a sharp 
 phase transition, but under questionable assumptions.
In Bonabeau \textit{et al}, strength is allowed to go
 to infinity with time. In Stauffer's model \cite{Stauffer},
when a moving average process is introduced to limit strength,
the transition is not sharp; Stauffer had to introduce
 some extra dynamics on the 
 $\beta$ discrimination parameter to get a sharp transition.
    The first motivation for the present research 
was to check the importance of the random "walkers"  dynamics
on the sharpness of the transition: would fixed agents, at the nodes of a social network, 
display a sharp transition towards some ''anti-ferromagnetic'' regime?

\section{The model and its numerical implementation}
  
   We use a square lattice with periodic boundary conditions.
Agents don't move and remain at lattice nodes.
At each time step, an agent $i$,
 randomly selected, updates its internal variable
 $h_i(t)$ through a random process. The agent interacts
with one of its four neighbours randomly chosen. He wins
with a probability $P$ related to its neighbour's strength
by a logit (or thermal) function according to:
\begin{equation}
  P = \frac{1}{1 + \exp (\beta (h_j(t) - h_i(t))}
\end{equation}
  As in thermal physics, a large $\beta$ coefficient
results in a nearly deterministic choice in favour of the stronger agent,
and in a nearly random choice when $\beta$ is small.
As a result of the interaction, the winner's strength
is increased by one, and the looser strength is decrease by one.
Suppose, e.g., that $i$ wins, the updating gives:
\begin{eqnarray}
  h_i(t)=(1-\gamma) h_i(t-1)+1 \quad ,\\
  h_j(t)=(1-\gamma) h_j(t-1)-1 \quad .
\end{eqnarray}

  These equations correspond to a moving average of 
past gains. Gains increase strength, but past gains are discounted 
at a rate $ 1-\gamma$. $1/\gamma$
is the characteristic time of the dynamics.
They imply one double update, since two sites $i$ and $j$ are
involved; one time step corresponds to one double update per site.

  We usually start simulations from a configuration
where all strengths are 0, and run them for a long time
to check the attractors of the dynamics.
  A typical time scale for the simulation is $10/\gamma$
updates per site, i.e. ten times the characteristic time.
This choice is also motivated by the equilibrium
amplitude of the strength of a constant winner:
\begin{eqnarray}
   h(eq)=(1-\gamma) h(eq)+1\\
    h(eq)= \frac{1}{\gamma}
\end{eqnarray}
  Times proportional to $1/\gamma$ are needed to allow
strength to saturate.

\section{Simulations results}

 Apart from the dimension of the lattice which might give rise to 
some size effects, two parameters control the dynamics,
$\beta$ the discrimination parameter and $\gamma$,
the memory parameter. Our simulations display the system
configuration at asymptotic time value or some kind of order parameter.

We have chosen the same order parameter as Bonabeau {\textit et al}
and followers,
namely  the standard deviation of the distribution of the probability 
to win, $\sigma$. We first checked at fixed $\gamma$ the evolution of $\sigma$
versus $\beta$, the discrimination coefficient (figure 1).

\begin{figure}
\centerline{\epsfxsize=120mm\epsfbox{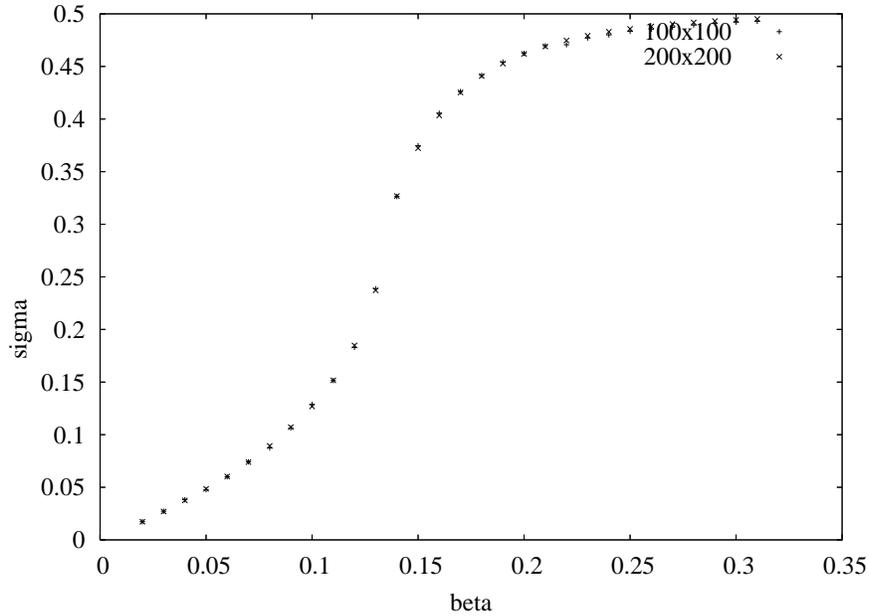}}
\caption{Standard deviation of the winning probability distribution
 as a function of $\beta$, observed at $\gamma = 0.1$ for networks of size 
$100 \times 100$ and $200 \times 200$. Simulated were 10 000 time steps.}
\end{figure}

   The observed crossover, starting at values of $\beta$
close to $\gamma$ is smooth and no size effect
is detectable between networks of size 10,000 and 40,0000
which could account for the observed smoothness.
  We also checked, for the same reason,
 the influence of convergence times which again
is not noticeable after a number of double updates per agent above  
$10/\gamma$ time steps (not represented here by a figure).

To check the influence of both $\beta$ and $\gamma$
it is interesting to use a reduced parameter.
$\beta/\gamma$ is a good candidate, by analogy
with ferromagnetism. The mean field theory 
of   ferromagnetism predicts a transition when
$\beta  = z \times J$, where $J$ is the coupling constant and  
$z$ the number of neighbours. We expect then 'something' to happen 
as a function of $\beta/\gamma$. Another model based on 
a logit choice and a moving average of a
 utility function based on past experience
is Weisbuch {\textit et al} model of Marseille fishmarket \cite{Weisbuch} which
also gives $\beta/\gamma$ as a reduced parameter.
$1/\gamma$ being the characteristic
strength involved in the present problem,
$\beta/\gamma$ is then a reasonable choice.

Figure 2 displays the standard deviation of the 
distribution of the probability 
to win, $\sigma$,  as a function of 
$\beta/\gamma$, 

\begin{figure}
\centerline{\epsfxsize=120mm\epsfbox{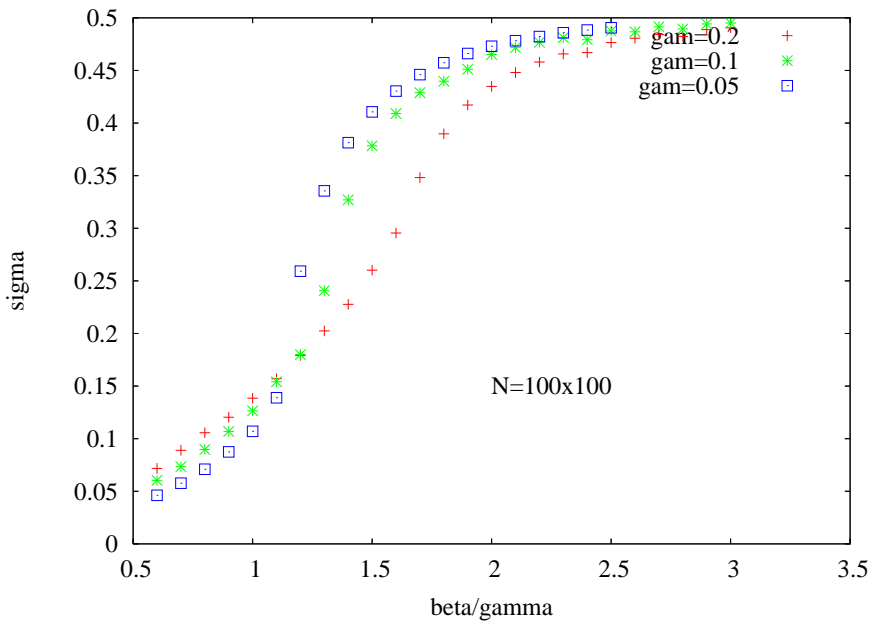}}
\caption{Standard deviation of the 
winning probability distribution as a function
 of $\beta/\gamma$, observed for networks
of size $100 \times 100$. $\gamma$ values are 0.05, 0.1 and 0.2.
Simulation times per site are  $10/\gamma$.}
\end{figure}
   
As we guessed, the onset of a smooth crossover occurs
slightly above $\beta/\gamma = 1$ where the three curves 
meet. And $\sigma$ saturates at 0.5, which corresponds to all probabilities
equally distributed between the two values 0 and 1. 
 But the function $\sigma$ vs. $\beta/\gamma$
is not universal: the slope of the line
near the crossover seems to increase with decreasing $\gamma$.
The Bonabeau {\textit et al} result, a vertical transition for 
infinite strength, is then consistent with ours for small $\gamma$.
The same is true for Stauffer's result: no sharp transition
for finite strength values. In other words, our fixed fighters 
results are consistent with those obtained for walking
fighters models.

   A similar behaviour is observed in figure 3 for the relative
 average amplitude of sites strength, another possible order parameter.
At large $\beta/\gamma$ values, strength amplitudes saturate to $1/\gamma$.
The curves cross around $\beta/\gamma=1$, but they don't collapse. 

\begin{figure}
\centerline{\epsfxsize=120mm\epsfbox{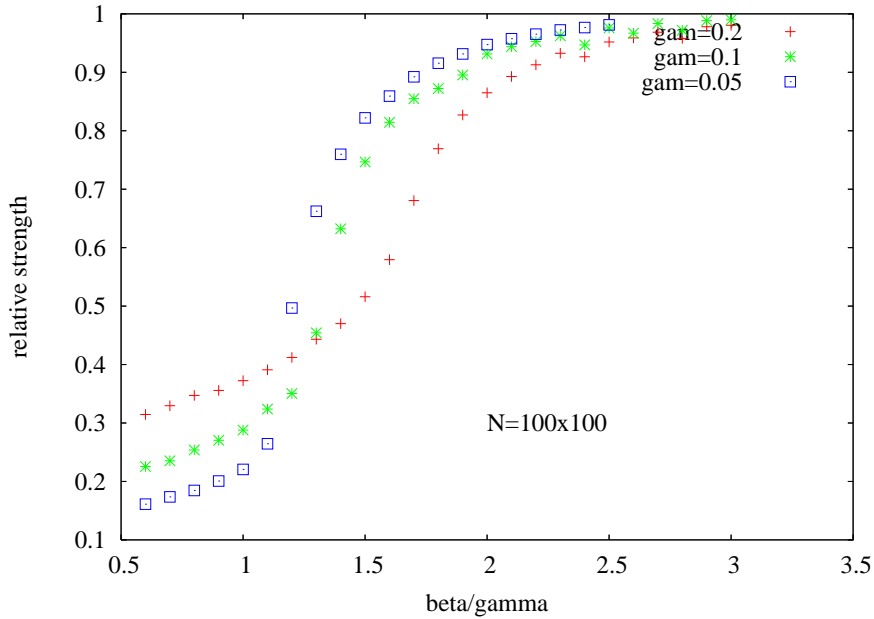}}
\caption{ Normalised averaged strength  amplitude $|h| \times \gamma $
 as a function of $\beta/\gamma$, observed for networks
of size $100 \times 100$. $\gamma$ values are 0.05, 0.1 and 0.2.
Simulation times per site are  $10/\gamma$.}
\end{figure}

  The histograms displayed in figure 4 provide a more direct  
representation of the state of the system at different $\beta$
values. Strengths vary between +10 and --10, respectively $\pm 1/\gamma$.
\begin{figure}
\centerline{\epsfxsize=120mm\epsfbox{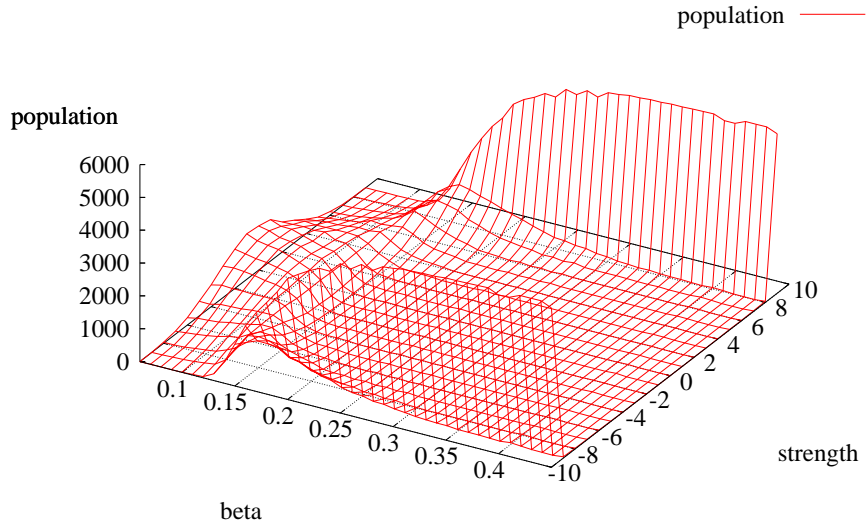}}
\caption{Strength distribution as a function of $\beta$, observed for networks
of size $100 \times 100$. $\gamma=0.1$. Simulation times were 100 000 time steps.}
\end{figure}
  In the low  $\beta$ region, say 
 $\beta \le \gamma$, the strength distribution is a Gaussian centered around
 0. By contrast, when $\beta \gg \gamma$, two narrow peaks are observed 
at $\pm 1/\gamma$. In between the transition is smooth. 

   Figure 5 is a set of $50 \times 50$ strength patterns obtained
after 10 000 time steps for different values of $\beta$
 below and above  $1/\gamma$.

\begin{figure}
\centerline{\epsfxsize=120mm\epsfbox{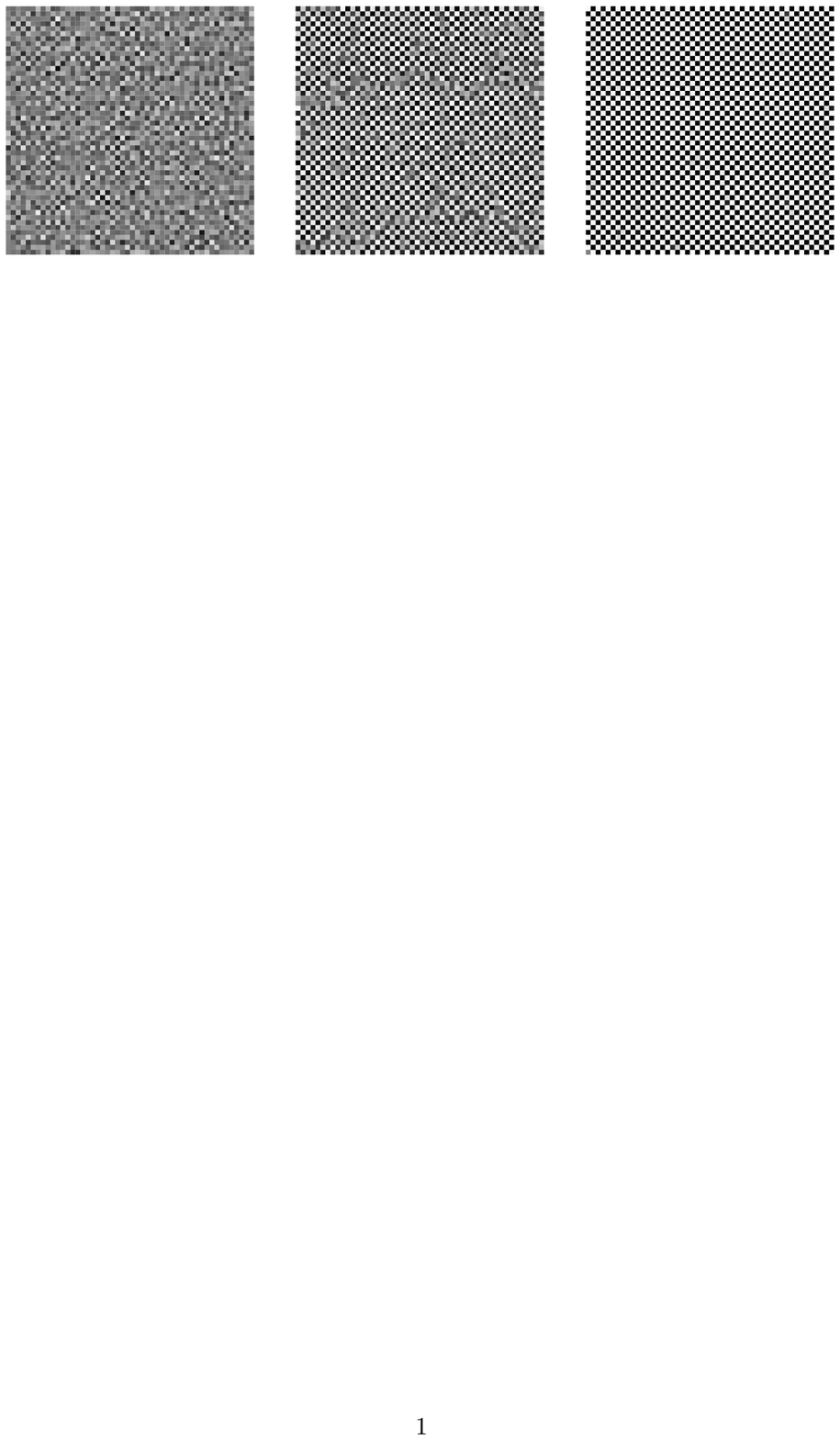}}
\caption{Strength patterns obtained on $50 \times 50$ lattices
after 10 000 time steps for $\beta=0.11$,
0.15 and 0.3; $\gamma=0.1$. Black corresponds to --10 strength,
grey to intermediate values, and white to 10 strength.
When $\beta$ increases from 0 to 0.3, 'anti-ferromagnetic' islands with 
checkerboard structure increase in size and invade the whole lattice} 
\end{figure}

   The checkerboard pattern obtained when $\beta=0.3$
is a clear picture of the analogous of an ''anti-ferromagnetic" configuration:
all sites have a maximum strength $1/\gamma$, alternately positive and
negative. For the intermediate values of $\beta$,
anti-ferromagnetic islands of size increasing with $\beta$
invade a sea of low-strength sites coloured in grey.
  At intermediate $\beta$ values,  the checkerboard islands do not grow in time
since the low strength agents at their boundary have one neighbour
within the checkerboard structure against 3 outside. They tend to relax towards
the average strength after their interactions with checkerboard sites.

  This intermediate regime can be compared to type II superconductors which
also  exhibit a mixed regime between two values of the external
magnetic field, where magnetisation increases continuously with the field.
In the mixed regime, 'normal' vortices regions are surrounded by
superconductive regions. 

  As observed in figure 6, the time evolution of the patterns in the anti-ferromagnetic
region implies two characteristic times: first,
ferromagnetic domains separated by low strength lines appear quite fast.
Two domains separated by a line are shifted by one site; in one dimension
the structure would be:

 \centerline { + -- + -- + -- + 0 -- + -- + -- + -- + --}

\noindent
(where + is positive, -- negative and 0 weak strength). The two
 antiferromagnetic domains are phase shifted by $\pi$ with respect to the lattice structure.

 The subsequent annealing process with coarsening of anti-ferro\-magnetic domains into a unique
domain is much longer.

   The online observation of the sites dynamics shows that two
factors account for the slow convergence
 dynamics and for the finite width of the crossover:
\begin{itemize}
\item The structuring process into antiferromagnetic patches
 begins early in time and when   $\beta/\gamma \simeq 1$.  But patches
growing from different regions of the lattice don't have any reason to be 
in phase. Whenever intermediate strength ($h \simeq 0$) agents interact 
against agents from the antiferromagnetic patches,
 their strength is increased (in absolute value) by interaction  
  with one side but it is decreased when they interact with the other side.
\item Since the strength difference between intermediate agents 
and agents inside patches is half of the strength difference inside
the patches, a higher $\beta$ value is necessary to maintain
long range order.
\end{itemize}

\begin{figure}
\centerline{\epsfxsize=120mm\epsfbox{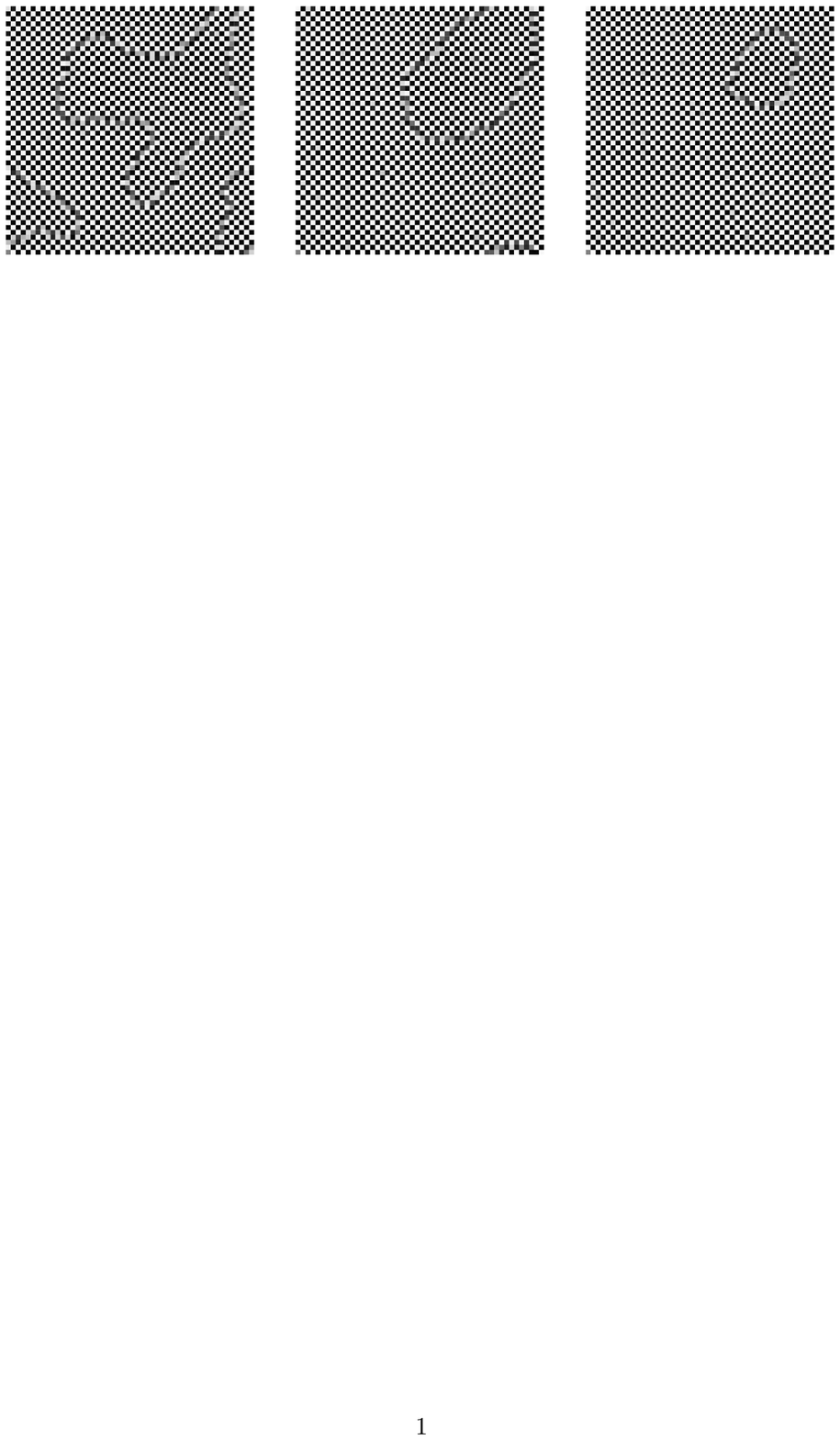}}
\caption{Time evolution of strength patterns obtained on $50 \times 50$ lattices
 for $\beta=0.3$ and $\gamma=0.1$, after 1000, 5000 and 8000 time steps.
The colour code is the same as above.
One first observes a fast evolution towards  a partial anti-ferromagnetic structure
with low strength line boundaries. Annealing of the boundaries
takes a longer time and is achieved after 10000 time steps as observed in the
previous figure.}
\end{figure}

\begin{figure}
\begin{center}
\includegraphics[angle=-90,scale=0.4]{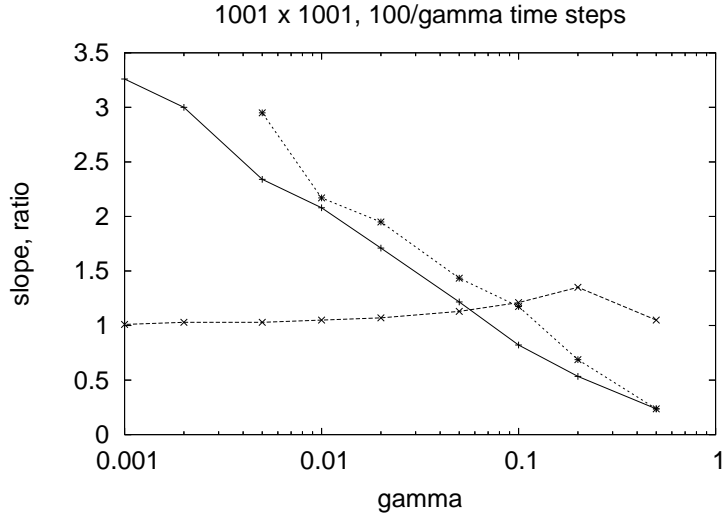}
\end{center}
\caption{Maximum slope (+,*), of the $\sigma$ versus 
$\beta/\gamma$ curves like figure 2, on $1000 \times 1000$ lattices for
$100/\gamma$ (+) and $1000/\gamma$ (*) time steps. The x symbols indicate the 
ratio $\beta/\gamma$ for which this slope is reached.
}
\end{figure}

Finally,  figure 7 returns to the question posed already by Fig.2: Would we get
for $\gamma \rightarrow 0$ a sharp phase transition in the sense that the slope
of $\sigma$ versus the ratio $\beta/\gamma$ would become infinite at some 
critical value of this ratio? For this purpose we simulated much larger 
lattices, for at least $100/\gamma$ time steps and $0.001\le\gamma <1$. We also
used regular updating, instead of the above random updating, and made minor
changes. There are long-time effects preventing complete equilibration, 
presumably due to reptation of domain boundaries. Nevertheless
this figure confirms our guess from Fig.2 about a sharp transition in the
limit $\gamma \rightarrow 0$. The critical ratio $\beta/\gamma$ seems to be 1.

\section{Conclusions} 

  Our first aim was to compare a fixed agents model of 'fighters' with  results 
obtained for 'walkers' models of Bonabeau {\textit et al}.
 We wanted to know whether the smoothness of the crossover 
observed when strength remains finite was due to the extra randomness 
introduced by walkers diffusion. We demonstrated that a fixed agent model 
also displays a smooth crossover.

 The observation of equilibrium patterns
shows that small anti-ferromagnetic domains
 start to develop in the intermediate 
mixed region, when $\beta \simeq \gamma$,
 but infinite long range order is only achieved 
for values of the discrimination parameter $\beta$ above 
the onset of the mixed region. The width of the mixed region
appears to increase with $\gamma$.

  Interpretations of the lattice in terms of social networks
show that the hierarchy is re-enforced by the fact each agent
interacts with the same neighborhood.
  On the other hand, social networks are
more random than lattices. They are not totally
random as Erd\"os-R\'enyi nets and display more betweenness
(small loops), than regular lattices. But the stability of the
 ordered regime obtained with square lattices containing 
only non-frustrated loops cannot be expected from 
more random structures: frustrated loops 
introduce glassy dynamics, with wider crossover
and more irreducible fluctuations in the ordered regime.
  
\vskip 1 cm  

{\bf Acknowledgments}

  The present research was inspired by the
GIACS summer school, Kazimierz Dolny, September 2006. We thank Sorin Solomon
for hospitality at Hebrew University where this work was completed.
We acknowledge fruitful discussions with Bernard Derrida 
and Jean Vannimenus.

\end{document}